\documentstyle[aps,preprint]{revtex}

\begin{document}

\draft

\preprint{Alberta-Thy-02-01; ~quant-ph/0101076}

\tightenlines

\title{Classical and Quantum Action-Phase Variables
for Time-Dependent Oscillators}

\author{Sang Pyo Kim\footnote{E-mail: spkim@phys.ualberta.ca}
and Don N. Page\footnote{E-mail:
don@phys.ualberta.ca}\footnote{Fellow, Cosmology and Gravity
Program of the Canadian Institute for Advanced Research}}

\address{Theoretical Physics Institute\\
Department of Physics\\ University of Alberta\\ Edmonton, Alberta,
Canada T6G 2J1}


\date{\today}

\maketitle

\begin{abstract}
For a time-dependent classical quadratic oscillator we introduce
pairs of real and complex invariants that are linear in position
and momentum. Each pair of invariants realize explicitly a
canonical transformation from the phase space to the invariant
space, in which the action-phase variables are defined.  We find
the action operator for the time-dependent quantum oscillator via
the classical-quantum correspondence. Candidate phase operators
conjugate to the action operator are discussed, but no
satisfactory ones are found.
\end{abstract}


\section{Introduction}

A time-independent classical oscillator is exactly solved when the
evolution of the position and momentum in the phase space is
completely known in terms of the initial data or constants of
motion. A constant of motion is a function of the initial data in
phase space, and the evolution of an oscillator is completely
determined by any two independent constants. For any linear
second-order system, such as a time-dependent classical
oscillator, two invariants (constants of motion) are readily
constructed from two independent classical solutions. The
time-dependent quantum oscillator can be exactly solved in terms
of the classical solutions, as the invariant operator, which was
first introduced by Lewis and Riesenfeld and is quadratic in
position and momentum, is known explicitly in terms of the
solutions of the corresponding classical oscillator \cite{lewis2}.
The invariant operator generates the Fock space of exact quantum
states of the Schr\"{o}dinger equation. Recently the
time-dependent quantum oscillator has attracted much attention
with the discovery of the geometric phase by Berry to study the
nonadiabatic geometric phase for various quantum states, such as
Gaussian, number, squeezed, or coherent states, which can be found
exactly \cite{jarzynski,song}.

In spite of intensive study of time-dependent oscillators, the
issue of action-phase (angle) variables remains not fully
exploited classically and quantum mechanically. In particular, the
quantum action-phase operators are still elusive and not
completely resolved since it was raised by Dirac \cite{dirac}. For
a time-independent classical oscillator the Hamiltonian itself,
being a constant of motion, turns out to be an action variable,
and its conjugate is the phase (angle) variable. The
transformation from the phase space to the action-phase space is a
canonical transformation. A generating function that yields the
action-phase variables can be found even for the time-dependent
classical oscillator \cite{arnold}. On the other hand, for a
time-independent quantum oscillator many different methods and
schemes have been introduced to define the action-phase operators
\cite{lynch}. Recently a candidate for a phase operator conjugate
to the Hamiltonian (action) operator has been proposed for the
time-independent quantum oscillator \cite{lewis3}.

The aim of this paper is to find the canonical transformations
from the phase space to the invariant spaces and to define the
action-phase variables for a time-dependent quadratic oscillator.
For that purpose we introduce pairs of classical invariants that
are linear in position and momentum and are expressed in terms of
the classical solutions, which realize explicitly the canonical
transformation from the phase space to the invariant spaces. We
find the most general three-parameter quadratic invariants formed
from the linear ones, which exhaust all quadratic invariants and
include the one by Lewis and Riesenfeld. In each invariant space
we define the action variable, quadratic in position and momentum,
and find its conjugate phase variable. Analogously by using pairs
of linear invariant operators that constitute the annihilation and
creation operators for the Fock space of exact quantum states
\cite{kim,kim-lee}, we find the action operator for the time-dependent
quantum oscillator. The three-parameter invariant operator is
transformed into the canonical form of a number operator through a
unitary (Bogoliubov) transformation. Similarly the invariant
operator obtained from one classical solution is transformed into
the one obtained from another classical solution through a unitary
transformation. A number eigenstate of one pair is a squeezed
state of the corresponding one of another pair. We extend the
current scheme for defining the phase operator to the
time-dependent quantum oscillator.

The organization of the paper is as follows. In Sec. II we find pairs
of linear invariants for the time-dependent classical oscillator
and show the canonical transformation from the phase space to the
invariant space. In each invariant space we define the
action-phase variables. In Sec. III we find the action operator
and define the phase operator conjugate to the action operator.

\section{Classical Action-Phase Variables}

We consider a time-dependent classical oscillator which is
described by the Hamiltonian
\begin{equation}
H(t) = \frac{X(t)}{2} p^2 + \frac{Y(t)}{2} (pq + qp) +
\frac{Z(t)}{2} q^2, \label{osc}
\end{equation}
where $X(t), Y(t)$ and $Z(t)$ depend explicitly on time. An
invariant for the Hamiltonian (\ref{osc}) is a constant of motion
and obeys the equation
\begin{equation}
\frac{d}{dt} I(t) = \frac{\partial}{\partial t} I(t) + \{I (t),
H(t) \}_{\rm PB} = 0, \label{inv eq}
\end{equation}
where $\{ ~,\}_{\rm PB}$ denotes a Poisson bracket. To find any
invariant it is necessary to find a pair of invariants, say, $I_1$
and $I_2$, which are linear in position and momentum and are
independent each other, because the invariants satisfying Eq.
(\ref{inv eq}) satisfy the multiplication property
\begin{equation}
\frac{\partial}{\partial t} (I_1 I_2) + \{I_1 I_2, H \}_{\rm PB} =
\Biggl[\frac{\partial}{\partial t} I_1 + \{I_1, H \}_{\rm PB}
\Biggr] I_2 + I_1 \Biggl[\frac{\partial}{\partial t} I_2 + \{I_2,
H \}_{\rm PB} \Biggr] = 0. \label{mul}
\end{equation}
Hence any analytical function of the invariants $I_1$ and $I_2$ is
also an invariant. For that purpose, we introduce two invariants 
(see Appendix A)
\begin{eqnarray}
a_1 (t) = u_1 (t) p - \frac{1}{X} [\dot{u}_1 (t)  - Y u_1 (t)] q,
\nonumber\\ a_2 (t) = u_2 (t) p - \frac{1}{X} [\dot{u}_2 (t) - Y
u_2 (t)] q, \label{inv}
\end{eqnarray}
where $u_1$ and $u_2$ are two independent real
solutions to the classical equation of motion
\begin{equation}
\frac{d}{dt} \Biggl(\frac{\dot{u}}{X} \Biggr) + \Biggl[ XZ - Y^2 +
\frac{\dot{X}Y - X \dot{Y}}{X}\Biggr] \Biggl(\frac{u}{X} \Biggr) =
0, \label{cl eq}
\end{equation}
where overdots denote $d/dt$. We require that these solutions are
normalized to satisfy the Wronskian condition
\begin{equation}
\frac{1}{X} (u_1 \dot{u}_2 - \dot{u}_1 u_2) = 1. \label{wr}
\end{equation}
Then $a_1$ and $a_2$ are canonical conjugates of each other,
satisfying
\begin{equation}
\{a_1 (t), a_2 (t)\}_{\rm PB} = 1. \label{pb}
\end{equation}
It should be remarked that there are as many pairs of independent
invariants as pairs of independent solutions. These pairs of
invariants constitute canonical conjugate pairs, satisfying Eq.
(\ref{pb}), if Eq. (\ref{wr}) is satisfied. As any two independent
solutions to Eq. (\ref{cl eq}), say $v_1$ and $v_2$, can be
expressed in terms of the original pair, $u_1$ and $u_2$, these
pairs are related to each other through a canonical
transformation. In quantum theory the canonical transformation
corresponds to a Bogoliubov transformation and has the meaning of
squeezing the quantum states, which will be discussed Sec. III.

For a time-independent oscillator with constant $X, Y$ and $Z$,
the Hamiltonian (\ref{osc}) is an action variable. For the
time-dependent oscillator of this paper, the Hamiltonian cannot
become an action variable, since $\partial H(t)/\partial t \neq
0$. Instead we look for the most general quadratic invariant of
the form
\begin{equation}
{\cal I} = \frac{1}{2} (B + A^* + A) a_1^2 + \frac{1}{2i} (A^* - A)
2 a_1 a_2 + \frac{1}{2} (B - A^* - A) a_2^2, \label{three par}
\end{equation}
where $A$ is a complex constant and $B$ is a real constant. In
fact, Eq. (\ref{three par}) describes real three-parameter
invariants that exhaust all real invariants that are quadratic in
position and momentum. Through a canonical transformation, which
is a rotation plus a scaling in the $(a_1, a_2)$ space, Eq.
(\ref{three par}) can be transformed into the canonical form
\begin{equation}
{\cal I} (t) = \frac{1}{2} [a_1^2 (t) + a_2^2 (t)]. \label{act}
\end{equation}
The canonical transformation has the physical meaning of squeezing
in quantum theory, which will be discussed in Sec. III. Constant
values of the action (\ref{act}) depict circles in the $(a_1,
a_2)$ plane with polar coordinates defined through
\begin{equation}
a_1 (t) = \sqrt{2{\cal I} (t)} \cos \theta (t), \quad a_2 =
\sqrt{2 {\cal I}(t)} \sin \theta (t).
\end{equation}
The phase variable, simply given by
\begin{equation}
\theta (t) = \tan^{-1} \Biggl[\frac{a_2 (t)}{a_1 (t)} \Biggr]
\label{ang}
\end{equation}
up to arbitrary additive constant, is an invariant, and is the
conjugate of the action (\ref{act}):
\begin{equation}
\{{\cal I } (t), \theta (t) \}_{\rm PB} = 1.
\end{equation}
It is then trivial to show that
\begin{equation}
\{{\cal I} (t), \tan \theta (t) \}_{\rm PB} =  \sec^2 \theta (t).
\end{equation}
The transformations from $(q, p)$ to $(a_1, a_2)$ and from $(a_1,
a_2)$ to $({\cal I}, \theta (t))$ are canonical, so their volume
elements are preserved:
\begin{equation}
dq \wedge dp = da_1 \wedge da_2 = d{\cal I} \wedge d\theta.
\end{equation}
The volume of the phase space with ${\cal I} \leq {\cal I}_0$ is
then
\begin{equation}
\int_{{\cal I} \leq {\cal I}_0} dq \wedge dp = \int_{{\cal I} \leq
{\cal I}_0} da_1 \wedge da_2 = \int_{{\cal I} \leq {\cal I}_0}
d{\cal I} \wedge d \theta = 2 \pi {\cal I}_0.
\end{equation}

To make manifest the correspondence between the classical and
quantum theory in Sec. III, we introduce another pair of
invariants
\begin{eqnarray}
a (t)  &=& (i) \Biggl\{u^*(t) p - \frac{1}{X(t)} [ \dot{u}^* (t) -
Y(t) u^* (t)] q \Biggr\}, \nonumber\\ a^* (t) &=& (-i) \Biggl\{u
(t) p - \frac{1}{X(t)} [ \dot{u} (t) - Y(t) u (t)] q \Biggr\},
\label{fun inv}
\end{eqnarray}
where $u$ is a complex solution to Eq. (\ref{cl eq}), normalized
to satisfy the Wronskian condition
\begin{equation}
\frac{1}{X} (u \dot{u}^* - u^* \dot{u})  = i. \label{wr2}
\end{equation}
$a$ and $a^*$ are the complex conjugates of each other and satisfy
the Poisson bracket
\begin{equation}
\{ a (t), a^* (t) \}_{\rm PB} = - i. \label{pb1}
\end{equation}
The factor $-i$ is a consequence of complex nature of $a$ and
$a^*$. If $u(t) = [iu_1 (t) + u_2 (t)]/\sqrt{2}$, the complex invariants
(\ref{fun inv}) are given by
\begin{equation}
a (t) = \frac{a_1 (t) + i a_2 (t)}{\sqrt{2}}, \quad a^* (t) =
\frac{a_1 (t) - i a_2 (t)}{\sqrt{2}}.
\end{equation}
Hence there is a canonical transformation from the phase space
$(q, p)$ to the invariant space $(a, a^*)$. The action variable
(\ref{act}) is now given by
\begin{eqnarray}
{\cal I} (t) &=& a^* (t) a (t) \nonumber\\ &=& u^* u \Biggl\{p +
\Biggl[\frac{Y}{X} - \frac{1}{X} \frac{d}{d t} \ln (u^* u)^{1/2}
\Biggr] q \Biggr\}^2 + \frac{q^2}{4 u^* u}. \label{act2}
\end{eqnarray}
A general quadratic invariant of $a$ and $a^*$ can be transformed
into the form (\ref{act2}) through a canonical transformation,
which has the physical interpretation of squeezing in quantum
theory, as will discussed in Sec. III.

As $a^*$ is the complex conjugate of $a$, the phase of $a$ can be
defined by
\begin{equation}
a (t) = \sqrt{{\cal I}(t)} e^{i \theta_a (t)}, \quad a^* (t) =
\sqrt{{\cal I}(t)} e^{- i \theta_a(t)}. \label{a pol}
\end{equation}
The polar form in Eq. (\ref{a pol}) leads to the phase variable in
the form
\begin{equation}
\theta_a (t) = \frac{i}{2} [ \ln a^* (t) - \ln a (t)].
\label{phase2}
\end{equation}
The equivalence between the phases (\ref{ang}) and (\ref{phase2})
can be shown using the identity
\begin{equation}
\theta (t) = \tan^{-1} \Biggl[\frac{a_2(t)}{a_1 (t)}\Biggr] =
\frac{i}{2} \ln \Biggl[\frac{i + (a_2(t)/a_1(t))}{i -
(a_2(t)/a_1(t))} \Biggr] = \theta_a (t).
\end{equation}
It follows that $\theta_a$ is the conjugate variable of ${\cal
I}$:
\begin{equation}
\{{\cal I}(t), \theta_a (t)\}_{\rm PB} = 1.
\end{equation}
After expressing $q$ and $p$ in (\ref{fun inv}) in terms of $a$
and $a^*$ by using (\ref{wr2}), and writing the complex solution
in a polar form
\begin{equation}
u (t) = \rho (t) e^{- i \theta_u (t)},
\end{equation}
we find
\begin{eqnarray}
q &=& u a+ u^* a^* = 2 \sqrt{\cal I} \rho \cos (\theta_a -
\theta_u), \nonumber\\ p &=& \frac{1}{X} (\dot{u} a + \dot{u}^*
a^*) - \frac{Y}{X} (u a+ u^* a^*) = \frac{2 \sqrt{\cal I}}{X}
(\dot{\rho} - Y \rho) \cos (\theta_a - \theta_u) +
\frac{\sqrt{\cal I}}{\rho} \sin(\theta_a - \theta_u).
\end{eqnarray}

There is also another phase variable from the action (\ref{act2}).
As Eq. (\ref{act2}) implies that constant ${\cal I}$ describes an
ellipse in the $(q, p)$ phase space, the phase variable can be
defined by \cite{song}
\begin{eqnarray}
\cos \vartheta &=& \sqrt{\frac{1}{4 u^* u {\cal I}}}q ,
\nonumber\\ \sin \vartheta &=& \sqrt{\frac{u^* u}{\cal I}}
\Biggl\{p + \Biggl[\frac{Y}{X} - \frac{1}{X} \frac{d}{dt} \ln (u^*
u)^{1/2} \Biggr] q \Biggr\}. \label{phase1}
\end{eqnarray}
The phase (\ref{phase1}) is the conjugate variable of the
invariant through the generating function
\begin{equation}
\vartheta = \frac{\partial S}{\partial {\cal I}},
\end{equation}
where
\begin{equation}
S({\cal I}, q; t) = \int_{q_0; {\cal I} = {\rm fixed}}^{q} p
({\cal I}, q; t) dq.
\end{equation}
Hence the two phase variables (\ref{phase1}) and (\ref{a pol}) are
related by
\begin{equation}
\vartheta = \theta_a - \theta_u.
\end{equation}
The canonical transformations from $(q, p)$ to the invariant space
of action-phase variables $({\cal I}, \theta)$ or $({\cal I},
\vartheta)$ preserve the phase volume
\begin{equation}
dq \wedge dp = d {\cal I} \wedge d\theta   =   d{\cal I} \wedge d
\vartheta.
\end{equation}
The area for ${\cal I}_0 \leq {\cal I}$ is
\begin{equation}
\int_{{\cal I} \leq {\cal I}_0} dq \wedge dp = \int_{{\cal I} \leq
{\cal I}_0} d{\cal I} \wedge d \theta   = \int_{{\cal I} \leq
{\cal I}_0} d{\cal I} \wedge d \vartheta   = 2 \pi {\cal I}_0.
\end{equation}

\section{Quantum Action-Phase Operators}

In quantum theory invariant operators obey the quantum
Liouville-von Neumann equation
\begin{equation}
i \hbar \frac{\partial}{\partial t} \hat{I} (t) + [\hat{I} (t),
\hat{H} (t)] = 0.
\end{equation}
Corresponding to the pair of classical invariants (\ref{fun inv}),
there are a pair of invariant operators \cite{kim,kim-lee}
\begin{eqnarray}
\hat{a} (t)  &=& \frac{i}{\sqrt{\hbar}} \Biggl[u^*(t) \hat{p} -
\frac{1}{X(t)} [ \dot{u}^* (t) - Y(t) u^* (t)] \hat{q} \Biggr],
\nonumber\\ \hat{a}^{\dagger} (t) &=& - \frac{i}{\sqrt{\hbar}}
\Biggl[u (t) \hat{p} - \frac{1}{X(t)} [ \dot{u} (t) - Y(t) u (t)]
\hat{q} \Biggr], \label{inv op}
\end{eqnarray}
where $u$ is a complex solution satisfying (\ref{wr}). Moreover,
the invariant operators satisfy the standard commutation relation
\begin{equation}
[\hat{a} (t), \hat{a}^{\dagger} (t) ] = 1. \label{com rel}
\end{equation}
There is the correspondence between the classical and quantum
theory via  $a \leftrightarrow \hat{a}$ and $a^* \leftrightarrow
\hat{a}^{\dagger}$, except for some operator ordering problems.
Just as the pair of classical invariants (\ref{fun inv}) are
fundamental in the sense that any analytic function $F(a, a^*)$ is
an invariant, the pair of invariant operators (\ref{inv op}) are
fundamental, since any analytic function $F(\hat{a}, \hat{a}^*)$
is also an invariant operator. This fact is guaranteed by the
multiplication property of solutions of the Liouville-von Neumann
equation,
\begin{equation}
i \hbar \frac{\partial}{\partial t} (\hat{I}_1 \hat{I}_2 ) +
[\hat{I}_1 \hat{I}_2, \hat{H}] =  \Biggl(i \hbar
\frac{\partial}{\partial t} \hat{I}_1 (t) + [\hat{I}_1, \hat{H} ]
\Biggr) \hat{I}_2 + \hat{I}_1 \Biggl( i \hbar
\frac{\partial}{\partial t} \hat{I}_2 + [\hat{I}_2, \hat{H}]
\Biggr) = 0.
\end{equation}
Hence the most general, quadratic, Hermitian, invariant operator
is spanned by $\hat{a}^{\dagger 2} (t)$, $\hat{a}^{\dagger} (t)
\hat{a} (t) + \hat{a} (t) \hat{a}^{\dagger} (t)$, and $\hat{a}^2
(t)$ with one complex and one real parameter, which corresponds to
the classical one (\ref{three par}).

The most general, quadratic, Hermitian invariant that can be
constructed from the pair (\ref{inv op}) has the form
\begin{equation}
\hat{I} = \frac{A}{2} \hat{a}^{\dagger2} (t) + \frac{B}{2}
[\hat{a}^{\dagger} (t) \hat{a} (t) + \hat{a} (t) \hat{a}^{\dagger}
(t)] + \frac{A^*}{2} \hat{a}^2 (t), \label{gen quad}
\end{equation}
where $A$ is a complex constant and $B$ is a real constant.
Another pair of invariant operators, say $\hat{b}(t)$ and
$\hat{b}^{\dagger} (t)$, obtained by replacing $u (t)$ by an
independent complex solution to Eq. (\ref{cl eq}), say $v(t)$, do
not change the general form (\ref{gen quad}), since these are
related to the original pair through the Bogoliuobov
transformation
\begin{eqnarray}
\hat{b} (t) &=& \alpha \hat{a} (t) + \beta \hat{a}^{\dagger} (t),
\nonumber\\ \hat{b}^{\dagger} (t) &=& \alpha^* \hat{a}^{\dagger}
(t) + \beta^* \hat{a} (t), \label{bog2}
\end{eqnarray}
where $\alpha$ and $\beta$ are determined by
\begin{eqnarray}
v(t) &=& \alpha^* u (t) - \beta^* u^* (t), \nonumber\\ v^* (t) &=&
\alpha u^* (t) - \beta u(t).
\end{eqnarray}
A unitary (Bogoliubov) transformation can be found,
\begin{eqnarray}
\hat{a} (t) &=& \cosh r \hat{\tilde{a}} (t) + \sinh r e^{i \delta}
\hat{\tilde{a}}^{\dagger} (t) \nonumber\\ &=& \hat{S} (r, \delta)
\hat{\tilde{a}} (t) \hat{S}^{\dagger} (r, \delta), \nonumber\\
\hat{a}^{\dagger} (t) &=& \cosh r \hat{\tilde{a}}^{\dagger} (t) +
\sinh r e^{- i \delta} \hat{\tilde{a}} (t) \nonumber\\ &=& \hat{S}
(r, \delta) \hat{\tilde{a}}^{\dagger} (t) \hat{S}^{\dagger} (r,
\delta), \label{bog1}
\end{eqnarray}
where the unitary operator is a squeezing operator
\begin{equation}
\hat{S} (z) = \exp \Biggl[\frac{1}{2} \{z^* \hat{\tilde{a}}^2 (t)
- z \hat{\tilde{a}}^{\dagger 2} (t)\} \Biggr], \quad z = r e^{\delta
+ \pi}, \label{sq op}
\end{equation}
which transforms (\ref{gen quad}) to a canonical form
\begin{equation}
\hat{\cal I} (t) = \frac{\tilde{B}}{2} [\hat{\tilde{a}}^{\dagger}
(t) \hat{\tilde{a}}(t) + \hat{\tilde{a}}
(t)\hat{\tilde{a}}^{\dagger} (t)].
\end{equation}
Here the squeezing parameter $r$, phase $\delta$, and the
coefficient $\tilde{B}$ are determined by
\begin{eqnarray}
e^{i \delta} \tanh r &=& \frac{1}{A^*} [- B \pm \sqrt{B^2 - A^*
A}], \nonumber\\ \tilde{B} &=& \frac{1}{2} [A \sinh 2r e^{-i
\delta} + B \cosh2r + A^* \sinh 2r e^{i\delta} ].
\end{eqnarray}
The physical meaning of the unitary transformation (\ref{bog2}) is
that the number state of $\hat{b}^{\dagger} (t) \hat{b} (t)$ is a
squeezed state of the corresponding one of $\hat{a}^{\dagger} (t)
\hat{a} (t)$. Similarly the number state of $\hat{a}^{\dagger} (t)
\hat{a} (t)$ is a squeezed state of the corresponding one of
$\hat{\tilde{a}}^{\dagger} (t) \hat{\tilde{a}} (t)$.

Without losing generality, we shall use the quantum action
operator
\begin{equation}
\hat{\cal I} (t) = \hbar \hat{a}^{\dagger} (t) \hat{a} (t) = \hbar
\hat{n} (t), \label{act op}
\end{equation}
which corresponds to the action variable (\ref{act}). 
By writing the complex solution to Eq. (\ref{cl eq}) as
\begin{equation}
u(t) = \sqrt{\frac{X(t)}{2}} \zeta (t)e^{- i \theta_{\zeta} (t)},
\end{equation}
Eq. (\ref{cl eq}) becomes a nonlinear auxiliary equation
\begin{equation}
\ddot{\zeta} + \Biggl[\Biggl(XZ - Y^2 + \frac{\dot{X}Y - X \dot{Y}}{X} \Biggr)
+ \Biggl(\frac{\ddot{X}}{2X} - \frac{3 \dot{X}^2}{4X^2} \Biggr) \Biggr] 
\zeta = \frac{1}{\zeta^3},
\end{equation}
where we used  $\dot{\theta}_{\zeta} = 1/\zeta^2$ from Eq. (\ref{wr2}).
Then the invariant operator (\ref{act op}) takes the form \cite{cervero}
\begin{equation}
\hat{\cal I} (t) = \frac{1}{2} \Biggl\{X \zeta^2 \Biggl[ \hat{p}
+ \Biggl(\frac{Y}{X} - \frac{\dot{\zeta}}{X \zeta} - 
\frac{\dot{X}}{2X^2} \Biggr) \hat{q}\Biggr]^2
+ \frac{1}{X\zeta^2} \hat{q}^2 \Biggr\}. 
\end{equation}
Another form of the complex solution,
\begin{equation}
u(t) = \frac{1}{\sqrt{2}} \xi (t)e^{- i \theta_{\xi} (t)},
\end{equation}
with $\dot{\theta}_{\xi} = X/\xi^2$, leads to the form \cite{gao}
\begin{equation}
\hat{\cal I} (t) = \frac{1}{2} \Biggl\{ \Biggl[ \xi\Biggl( \hat{p}
+ \frac{Y}{X} \hat{q} \Biggr) - \frac{\dot{\xi}}{X} \hat{q} \Biggr]^2
+ \frac{1}{\xi^2} \hat{q}^2 \Biggr\}, 
\end{equation}
where $\xi$ satisfies the auxiliary equation
\begin{equation}
\ddot{\xi} - \frac{\dot{X}}{X} \xi + \Biggl(XZ - Y^2 + \frac{\dot{X}Y - X \dot{Y}}{X} \Biggr) \xi = \frac{X^2}{\xi^3}.
\end{equation}
The invariant operator (\ref{act op}) is nothing but a number
operator. Each eigenstate of $\hat{\cal N}$ is a number state
\begin{equation}
\hat{n} (t) \vert n, t \rangle = n \vert n, t \rangle. \label{num
st}
\end{equation}
The wave function for the number state defined by Eq. (\ref{num
st}), up to an arbitrary constant phase factor, is (see Appendix B)
\begin{eqnarray}
\Psi_{n} (q, t) = \frac{1}{\sqrt{(2\hbar)^{n} n!}}
\Biggl(\frac{1}{2 \pi \hbar u^* u} \Biggr)^{1/4}  \Biggl(
\frac{u}{\sqrt{u^*u}} \Biggr)^{(2n +1)/2} H_{n}
\Biggl(\frac{q}{\sqrt{2 \hbar u^* u}} \Biggr) \exp
\Biggl[\frac{i}{2\hbar X} \Biggl(\frac{\dot{u}^*}{u^*}  - Y
\Biggr) q^2 \Biggr], \label{osc wav}
\end{eqnarray}
where $H_{n}$ is the Hermite polynomial. It satisfies the
time-dependent Schr\"{o}dinger equation. In the sense that one
finds the Fock space of exact quantum states of the
Schr\"{o}dinger equation, even the time-dependent quantum
oscillator (\ref{osc}) is exactly solvable in terms of the complex
classical solution $u(t)$.

Finally we turn to the phase operator conjugate to the action
operator (\ref{act op}). A phase operator should satisfy the
commutation relation
\begin{equation}
[\hat{\cal I}(t), \hat{\theta} (t)] = i \hbar, \label{com rel2}
\end{equation}
where $\theta$ is restricted to $(0, 2 \pi)$. The commutation
relation leads to the Lerner criterion
\begin{equation}
[e^{i \hat{\theta} (t)}, \hat{\cal I} (t)] =  \hbar e^{i
\hat{\theta} (t)}. \label{ler}
\end{equation}
There are many schemes to define the phase operator, and each
method has its own advantage and disadvantage \cite{lynch}. As
there is a formal similarity between time-independent and
time-dependent quantum oscillators, we follow straightforwardly
the various well-known definitions of phase operator. First, a
naive definition from (\ref{a pol}) according to the
correspondence principle would be
\begin{equation}
\hat{a} (t) = e^{i \hat{\theta}_{\rm D} (t)} \hat{n}^{1/2} (t),
\quad \hat{a}^{\dagger} (t) = \hat{n}^{1/2} (t) e^{- i
\hat{\theta}_{\rm D} (t)}, \label{dirac ph}
\end{equation}
which is the procedure taken by Dirac \cite{dirac}. The
$\hat{\theta}_{\rm D}$ formally satisfies the Lerner criterion (\ref{ler}),
but only if one ignores the fact that to get Eq. (\ref{ler})
formally, one must multiply Eq. (\ref{com rel}) by the singular operator
$\hat{n}^{-1/2}$. Furthermore, $\hat{\theta}_{\rm D} (t)$  
is not a Hermitian operator, so $e^{i
\hat{\theta}_{\rm D}(t)}$ is not unitary \cite{lynch}. Second, the
phase operator by Susskind and Glogower \cite{susskind} is defined
as
\begin{equation}
\hat{E} (t) \equiv [\hat{n} (t) + 1]^{-1/2} \hat{a} (t), \quad
\hat{E}^{\dagger} (t) \equiv \hat{a}^{\dagger} (t) [\hat{n} (t) +
1]^{-1/2}.
\end{equation}
$\hat{E}$ and $\hat{E}^{\dagger}$ are the analogs of $e^{i
\hat{\theta}}$ and $e^{-i \hat{\theta}}$, which satisfy the Lerner
criterion (\ref{ler}). The analogs of $\cos \hat{\theta}$ and
$\sin \hat{\theta}$ can be defined similarly. $\hat{E}$ and
$\hat{E}^{\dagger}$ have the equivalent number state
representation
\begin{equation}
\hat{E} (t) = \sum_{n = 0}^{\infty} \vert n, t \rangle \langle
n+1, t \vert, \quad \hat{E}^{\dagger} (t) = \sum_{n = 0}^{\infty}
\vert n + 1, t \rangle \langle n, t \vert. \label{susskind ph}
\end{equation}
However, $\hat{E}$ is a one-sided unitary operator,
\begin{equation}
\hat{E} (t) \hat{E}^{\dagger} (t) = 1, \quad \hat{E}^{\dagger} (t)
\hat{E} (t) = 1 - \vert 0, t \rangle \langle 0, t \vert.
\end{equation}
This is somewhat related with the ambiguity in defining a phase
for $a (t) = 0$ in classical theory or for the vacuum state with 
$\hat{a} (t) \vert 0, t \rangle = 0$ in
quantum theory. Third, the phase operator defined by Pegg and
Barnett in a finite dimensional subspace of Fock space\cite{pegg},
is now given by
\begin{equation}
\hat{\theta}_{\rm PB} (t) = \sum_{m = 0}^{s} \theta_m \vert
\theta_m, t \rangle \langle \theta_m, t \vert, \label{pegg ph}
\end{equation}
where
\begin{eqnarray}
\vert \theta_m, t \rangle &=& \frac{1}{\sqrt{s+1}} \sum_{n =
0}^{s} e^{i n \theta_m} \vert n, t \rangle, \nonumber\\ \theta_m
&=& \theta_0 + \frac{2m\pi}{s+1}, \quad (\theta_0 = {\rm
constant}).
\end{eqnarray}
Fourth, the phase operator corresponding to the classical phase
variable (\ref{phase2}) is naively given by
\begin{equation}
\hat{\theta}_a (t) = \frac{i}{2} [ \ln
\hat{a}^{\dagger} (t) - \ln \hat{a} (t) ]. \label{lewis ph}
\end{equation}
The phase operator (\ref{lewis ph}) formally satisfies the
commutation relation (\ref{com rel2}) and is a generalization of
the time-independent one in Ref. \cite{lewis3} to the
time-dependent one. However, it is not really well-defined, since
$\hat{a}$ has a zero eigenstate, so that its logarithm is a
divergent operator.

A few comments are in order. At present there is no consistent
definition of a phase operator for a quantum oscillator free from all
conceptual and technical problems \cite{lynch}. This dilemma is
somewhat rooted in the quantization of phase variable according to
Eq. (\ref{com rel2}):
\begin{equation}
\hat{\cal I} = i \hbar \frac{\partial}{\partial \theta}.
\label{act dif}
\end{equation}
The eigenfunctions of Eq. (\ref{act dif}), consistent with the
periodicity of $\theta$, are given by
\begin{equation}
\Psi_n (\theta) = \frac{1}{\sqrt{2 \pi}} e^{- i n \theta}, \quad (
n = {\rm all ~ integers}). \label{ph wav}
\end{equation}
The wave functions (\ref{ph wav}) form a basis of the Hilbert
space on a circle and make the closure complete when $n$ runs for
all integers. Denoting the wave functions (\ref{ph wav}) by $\vert
n, \theta \rangle $, the following phase operator \cite{newton}
\begin{equation}
e^{i \hat{\theta}} = \sum_{n = - \infty}^{\infty} \vert n, \theta
\rangle \langle n + 1, \theta \vert, \quad e^{- i \hat{\theta}} =
\sum_{n = - \infty}^{\infty} \vert n + 1, \theta \rangle \langle
n, \theta \vert \label{ext ph}
\end{equation}
becomes unitary
\begin{equation}
e^{i \hat{\theta}} e^{-i \hat{\theta}} = e^{-i \hat{\theta}} e^{i
\hat{\theta}} = 1.
\end{equation}
However, compatibility with Eqs. (\ref{num st}) restricts $n$ to
all non-negative integers. As Eq. (\ref{osc wav}) and (\ref{ph
wav}) are the coordinate and phase-representations of the action
eigenstate (\ref{num st}), the phase operator (\ref{ext ph}) is
the same obtained by extending Eq. (\ref{susskind ph}) over all
integers including negative ones. But there is no physical meaning
to the wave functions of oscillator with negative eigenvalues.

\section{Conclusion}

In summary, we found pairs of real and complex invariants for a
 time-dependent classical oscillator, which are linear in position
and momentum and transform canonically the phase space to
invariant spaces. An action was defined that is quadratic in each
invariant space, and its conjugate phase (angle) variable was
found. The relation among different phase variables was exploited.
Analogously, pairs of linear complex invariant operators were
found for the time-dependent quantum oscillator. The action
operator, which is the quantum analog of the classical action, may
be used to construct a Fock space of exact quantum number
eigenstates. As the invariant operators act as the annihilation
and creation operators, we followed the procedure for
time-independent oscillator to define the phase operator conjugate
to the action operator for the time-dependent oscillator.

The action-phase variables were defined for the time-dependent
classical oscillator exactly as for the time-independent
oscillator. However any known scheme to define a consistent phase
operator, even for the time-independent oscillator, confronts some
technical difficulty \cite{lynch}. The same degree of difficulty
remains in defining the phase operator for the time-dependent
oscillator. In this paper we did not intend to resolve all the
puzzling issues of phase operator but to extend the current
methods to the time-dependent oscillator. One interesting
possibility will be to make use of the Wigner function of the
exact quantum state. The Wigner function can be expressed in the
invariant space, whose integral along each ray of constant phase
may lead to a phase pseudo-probability distribution
\cite{kim-page}. Then the phase probability distribution can be
used to define an ensemble average of any physical function of the
phase variable in quantum theory.

\acknowledgements This work was supported by the Natural Sciences
and Engineering Research Council of Canada. The work of SPK was
also supported in part by the Korea Research Foundation under
Grant No. 2000-015-DP0080.

\appendix

\section{Two Linear Invariants}

In this appendix we derive two linear invariants given by Eq. 
(\ref{inv}) or (\ref{fun inv}).
The Hamiltonian (\ref{osc}) can be written as
\begin{equation}
H (t) = \frac{1}{2} P^2 + \frac{\omega^2 (t)}{2} Q^2,
\end{equation}
where $P$ and $Q$ are new canonical variables
\begin{equation}
Q = \frac{1}{\sqrt{X(t)}}q, \quad P = \sqrt{X(t)} \Biggl(p 
+ \frac{Y(t)}{X(t)}q\Biggr),
\end{equation}
and
\begin{equation}
\omega^2 (t) = X (t) Z(t) - Y^2 (t).
\end{equation}
The $Q$ and $P$ are canonical conjugates  of each other, satisfying
\begin{equation}
\{ Q, P \}_{\rm PB} = 1.
\end{equation} 
By introducing another pair of new canonical variables $a_1(t)$ 
and $a_2(t)$, satisfying
\begin{equation}
\{a_1 (t), a_2 (t)\}_{\rm PB} = 1,
\end{equation} 
we can express $Q$ and $P$ in terms of $a_1(t)$ and $a_2 (t)$
\begin{eqnarray}
Q = \frac{1}{\sqrt{X(t)}} [u_2 (t) a_1 (t) - u_1 (t) a_2 (t)], \nonumber\\
P = \frac{1}{\sqrt{X(t)}}[\dot{u}_2 (t) a_1 (t) - \dot{u}_1 (t) a_2 (t)],
\end{eqnarray}
and vice versa
\begin{eqnarray}
a_1 (t) = \frac{u_1 (t)}{\sqrt{X(t)}} P - \frac{\dot{u}_1 (t)}{\sqrt{X(t)}} Q, \nonumber\\
a_2 (t) = \frac{u_2 (t)}{\sqrt{X(t)}} P - \frac{\dot{u}_2 (t)}{\sqrt{X(t)}} Q.
\end{eqnarray}
Then the Hamiltonian has the form
\begin{equation}
H (t) =  - 2 H_0 (t) a_1 (t) a_2 (t) +  H_1 (t) a_1^2 (t) + H_2 (t) a_2^2 (t),
\end{equation}
where
\begin{eqnarray}
H_0 (t) &=& \frac{1}{2X (t)} [\dot{u}_1 (t) \dot{u}_2 (t) 
+ \omega^2 (t) u_1 (t) u_2 (t)], \nonumber\\
H_1 (t) &=& \frac{1}{2X (t)} [ \dot{u}_2^2 (t) + \omega^2 (t) u_2^2 (t)], \nonumber\\ 
H_2 (t) &=& \frac{1}{2X (t)} [\dot{u}_1^2 (t) + \omega^2 (t) u_1^2 (t)]. 
\end{eqnarray}
Finally we require $a_1 (t)$ and $a_2 (t)$ each to satisfy the 
invariant equation (\ref{inv eq}):
\begin{eqnarray}
0 &=& \frac{\partial}{\partial t} a_1 (t) + \{a_1 (t), H(t) \}_{\rm PB} \nonumber\\ &=&
\Biggl[u_2 \Biggl\{\frac{d}{dt} \Biggl(\frac{\dot{u}_1}{X} \Biggr) + \Biggl( XZ - Y^2 +
\frac{\dot{X}Y - X \dot{Y}}{X}\Biggr) \Biggl(\frac{u_1}{X} \Biggr) \Biggr\} \Biggr]a_1 (t) \nonumber\\
&& + \Biggl[u_1 \Biggl\{\frac{d}{dt} \Biggl(\frac{\dot{u}_1}{X} \Biggr) + \Biggl( XZ - Y^2 +
\frac{\dot{X}Y - X \dot{Y}}{X}\Biggr) \Biggl(\frac{u_1}{X} \Biggr) \Biggr\}\Biggr] a_2 (t), \nonumber\\
0 &=&  \frac{\partial}{\partial t} a_2 (t) + \{a_2 (t), H(t) \}_{\rm PB} \nonumber\\ &=&  \Biggl[u_1 \Biggl\{\frac{d}{dt} \Biggl(\frac{\dot{u}_2}{X} \Biggr) + \Biggl( XZ - Y^2 +
\frac{\dot{X}Y - X \dot{Y}}{X}\Biggr) \Biggl(\frac{u_2}{X} \Biggr) \Biggr\} \Biggr] a_1 (t) \nonumber\\
&&+ \Biggl[u_2 \Biggl\{\frac{d}{dt} \Biggl(\frac{\dot{u}_2}{X} \Biggr) + \Biggl( XZ - Y^2 +
\frac{\dot{X}Y - X \dot{Y}}{X}\Biggr) \Biggl(\frac{u_2}{X} \Biggr) \Biggr\}\Biggr] a_2 (t).
\end{eqnarray}
Therefore, $a_1 (t)$ and $a_2 (t)$ satisfy Eq. (\ref{inv eq}) provided that $u_1(t)$ and $u_2 (t)$ satisfy the classical equation of motion (\ref{cl eq}).

\section{Wave Functions}

In this appendix we derive the exact wave functions (\ref{osc wav}) for the 
time-dependent oscillator (\ref{osc}) that satisfy the time-dependent 
Schr\"{o}dinger equation
\begin{equation}
i \hbar \frac{\partial}{\partial t} \Psi (q, t) = \hat{H} (t) \Psi (q, t).
\label{time sch}
\end{equation}
First we note that the operators $\hat{a} (t)$ and $\hat{a}^{\dagger} (t)$ 
in Eq. (\ref{inv op}) are time-dependent 
analogs of time-independent annihilation and creation operators. In fact, 
in the time-independent case with constant $X_0$, $Y_0$, and $Z_0$, by 
choosing the complex solution to Eq. (\ref{cl eq}) given by
\begin{equation}
u (t) = \frac{1}{\sqrt{2 \omega_0}} e^{- i \omega_0 t}, \quad \omega_0^2 = 
X_0 Z_0 - Y^2_0,
\end{equation}  
it can be shown that $\hat{a} (t) = e^{- i \omega_0 t} \hat{a}_0 $ and 
$\hat{a}^{\dagger} (t) = e^{i \omega_0 t} \hat{a}_0^{\dagger}$, where
$\hat{a}_0$ and $\hat{a}^{\dagger}_0$ are time-independent annihilation 
and creation operators. The time-dependent phase factors are necessary 
to satisfy Eq. (\ref{time sch}). Hence $\hat{n}(t) 
= \hat{n}_0$ and number states are the same in both representations. 
We follow exactly the same procedure of the time-indepenent oscillator 
to obtain the wave functions for the time-dependent oscillator. 
The ground state wave function is annihilated by $\hat{a} (t)$
\begin{equation}
\hat{a} (t) \Psi_0 (q, t) = 0,
\end{equation}
so we obtain 
\begin{equation} 
\Psi_{0} (q, t) =  \Biggl(\frac{1}{2 \pi \hbar u^* u} \Biggr)^{1/4}  \Biggl(
\frac{u}{\sqrt{u^*u}} \Biggr)^{1/2} \exp
\Biggl[\frac{i}{2\hbar X} \Biggl(\frac{\dot{u}^*}{u^*}  - Y
\Biggr) q^2 \Biggr]. \label{gr wav}
\end{equation}
Here we chose the time-dependent phase factor $(u/\sqrt{u^* u})^{1/2}$ so that
Eq. (\ref{gr wav}) satisfies Eq. (\ref{time sch}).
The $n$th wave function for the $n$th excited state is obtained by applying 
$\hat{a}^{\dagger} (t)$ $n$ times: 
\begin{eqnarray}
\Psi_{n} (q, t) &=&  \frac{1}{\sqrt{n!}} \Biggl(\hat{a}^{\dagger} (t) \Biggr)^n
\Psi_0 (q, t) \nonumber\\
&=& \frac{1}{\sqrt{(2\hbar)^{n} n!}}
\Biggl(\frac{1}{2 \pi \hbar u^* u} \Biggr)^{1/4}  \Biggl(
\frac{u}{\sqrt{u^*u}} \Biggr)^{(2n +1)/2} H_{n}
\Biggl(\frac{q}{\sqrt{2 \hbar u^* u}} \Biggr) \exp
\Biggl[\frac{i}{2\hbar X} \Biggl(\frac{\dot{u}^*}{u^*}  - Y
\Biggr) q^2 \Biggr]. \label{app wav}
\end{eqnarray}
It should be remarked that  the wave function 
(\ref{app wav}) satisfies not only Eq. (\ref{time sch}) but also the relation
\begin{equation}
\int dq \Psi^*_n (q, t) \Biggl(i \hbar \frac{\partial}{\partial t} \Biggr)
\Psi_n (q, t) = \int dq \Psi^*_n (q, t) \hat{H} (t) \Psi_n (q, t). 
\label{geo ph}
\end{equation}
Hence Eq. (\ref{geo ph}) implies that the geometric phase factor 
equals to the dynamical phase factor.
In the special case of $Y = 0$, the wave function (\ref{app wav}) 
reduces, up to 
the phase factor $(u/\sqrt{u^* u})^{1/2}$, to Eq. (3.9) 
of Ref. \cite{kim-lee}.

\end{document}